\documentclass[a4paper,11pt]{article}
\usepackage{pos}
\usepackage{multirow}
\usepackage{amsmath}
\usepackage{lineno}

\title{VV Resummation To NNLO+NNLL At the LHC}

\author[c]{Pulak Banerjee}
\author[b]{Chinmoy Dey}
\author[a]{M.C.Kumar}
\author*[a]{Vaibhav Pandey}

\affiliation[a]{Indian Institute of Technology Guwahati,\\ North Guwahati, Assam, India-781039}
\affiliation[b]{Theoretical Physics Division, Physical Research Laboratory,\\ Navrangpura, Ahmedabad 380009, India}
\affiliation[c]{Department of Physics, School of Advanced Sciences,\\ Vellore Institute of Technology (VIT), Chennai Campus, \\Chennai 600127, Tamil Nadu, India.}

\emailAdd{pulak.banerjee@vit.ac.in}
\emailAdd{chinmoy@prl.res.in}
\emailAdd{mckumar@iitg.ac.in}
\emailAdd{vphiitg@iitg.ac.in }

\abstract{
We present the resummed predictions for a vector boson pair production at the LHC. We have performed threshold resummation to next-to-next-to-leading logarithmic (NNLL) accuracy, and then matched them to next-to-next-to-leading order (NNLO) QCD results. After resummation, we observe a reduction in the scale uncertainties arising from unphysical renormalization and factorization scales. We find that the resummed corrections add a few per cent to the fixed order results for both ZZ and WW production.
}

\FullConference{17th International Symposium on Radiative Corrections: Applications of Quantum Field Theory to Phenomenology (RADCOR2025)\\
5-10 October 2025\\
Puri, India\\}

\begin{document}
\maketitle

\section{Introduction}

Vector-boson pair (VV, with V = W, Z) production at hadron colliders is a cornerstone process for testing the electroweak sector of the Standard Model and for constraining anomalous gauge couplings, while simultaneously constituting a major background to Higgs and many beyond-the-Standard-Model searches.
ZZ production is a crucial process for testing the electroweak symmetry-breaking mechanism.
WW production, on the other hand, is sensitive to the charged gauge boson self-interactions and is an important background for Higgs boson measurements, particularly in the $H \to WW^*$ decay channel.
Achieving per cent-level theoretical precision for WW and ZZ production cross-sections and distributions is therefore essential for fully exploiting current and future LHC data. 
It becomes crucial to include higher-order QCD corrections in theoretical predictions for these processes to achieve a precise comparison with experimental measurements.

Theoretical advances over the past few years have led to significant improvements in the precision of predictions for these processes.
The next-to-leading order (NLO) QCD corrections to on-shell WW and ZZ production were computed in the early 1990s \cite{Ohnemus:1990za, Frixione:1993yp}. 
The next-to-next-to-leading order (NNLO) QCD corrections were computed in the mid-2010s \cite{Gehrmann:2014fva, Cascioli:2014yka}, and the NLO electroweak (EW) corrections were obtained around the same time \cite{Grazzini:2019jkl}.
Using the SCET formalism, the resummation of soft-gluon effects in the threshold limit for these processes has been performed at next-to-next-to-leading logarithmic (NNLL), matched with NLO fixed-order results in \cite{Dawson:2013lya, Wang:2014mqt}.

In this paper, we perform threshold resummation for the on-shell production of WW and ZZ pairs at the LHC to NNLL accuracy, matched to fixed-order NNLO results. In section 2, we briefly outline the theoretical framework for the resummation. In section 3, we present the numerical results for the invariant mass distribution and the total cross-section, along with a study of the impact of resummation on the theoretical uncertainties.

\section{Theoritical framework}
In this section, we will briefly outline the theoretical framework that is used for this study. The computational framework we follow is the same as in \cite{Banerjee:2024cpr, Banerjee:2025tbo}.

The hadronic cross-section for VV production can be written as,
\begin{align}\label{eq:had-xsect}
 \frac{d\,\sigma}{d\, Q}  =
\sum_{a,b= \{q, \bar{q}, g\}}\int_0^1 dx_1\int_0^1 dx_2 \,\,f_{a}(x_1,\mu_F^2)\, f_{b}(x_2,\mu_F^2) \int_0^1 dz~ \delta \left(\tau-z x_1 x_2 \right) \frac{d\,\hat\sigma_{ab}}{d\, Q} \,,
\end{align}
where $Q$ is the invariant mass of the final state.
The hadronic and partonic threshold variables $\tau$
and $z$ are defined as
\begin{align}
\tau=\frac{Q^2}{S}, \qquad z= \frac{Q^2}{s} \,,
\end{align}
where $S$ and $s$ are the hadronic and partonic centre of mass energies, respectively. Thus, $\tau$ and $z$ are related by $\tau = x_1 x_2 z$.

The leading order parton level processes which are under study are,
\begin{align}\label{eq:parton}
  & q(p_1) + \bar{q}(p_2) \rightarrow Z(p_3) + Z(p_4).
  & q(p_1) + \bar{q}(p_2) \rightarrow W^{+}(p_3) + W^{-}(p_4).
\end{align} 

At higher orders in QCD, the partonic cross section receives contributions from both virtual loop corrections and real emissions. 
In the threshold region, corresponding to the limit $z \to 1$, nearly all the available energy in the partonic centre-of-mass frame is used to produce the final-state. 
As a result, the phase space for additional parton radiation becomes highly restricted, and soft-gluon effects play a dominant role. The partonic cross section, expressed in terms of $z$, can be written as:
\begin{align}
\frac{d\,\hat \sigma_{ab}}{d\, Q} =
	\frac{d\,\hat \sigma^{(0)}_{ab}}{d\, Q}\left(
	\Delta_{ab}^{\rm sv}(z,\mu_F^2)
	+ \Delta_{ab}^{\rm reg}(z,\mu_F^2)
\right) \,.
\end{align}

The contributions $\Delta_{ab}^{\rm sv}$ are universal in nature and depend only on the identity of the initial-state partons. 
The SV term collects all singular contributions arising in the threshold limit $z \to 1$, including the associated threshold logarithms. 
The remaining terms, denoted by $\Delta_{ab}^{\rm reg}$, are process dependent and contain logarithmic contributions away from threshold, as well as terms that remain finite as $z \to 1$. 
It is worth noting that only quark-antiquark and gluon-gluon initiated channels contribute to the SV component in the processes under study.

The singular part of the SV component is universal and gets contributions from such as virtual form factors~\cite{Moch:2005tm,
Moch:2005id,
Baikov:2009bg,
Gehrmann:2010ue,
Gehrmann:2014vha},
mass factorization kernels~\cite{Moch:2004pa, Vogt:2004mw, Blumlein:2021enk}
and soft-gluon emissions~\cite{Sudakov:1954sw,
Mueller:1979ih,
Collins:1980ih,
Sen:1981sd,
Sterman:1986aj,
Catani:1989ne,
Catani:1990rp,
Kidonakis:1997gm,
Kidonakis:2003tx,
Ravindran:2005vv,
Ravindran:2006cg,
Moch:2005ba,
Laenen:2005uz,
Kidonakis:2005kz,
Idilbi:2006dg}.

The process-dependent virtual contributions in the threshold limit appear as terms proportional to $\delta(1-z)$, while the universal contributions are expressed through plus distributions, ${\cal D}_i = [\ln^i(1-z)/(1-z)]_+$. These terms can be resummed to all orders in perturbation theory by working in Mellin space, where the convolutions simplify to ordinary products.

In Mellin space, the resummed SV cross-section can be written as:
\begin{align}\label{eq:resum-partonic}
	\hat{\sigma}_N^{\text{N}^n\text{LL}} =
	\int_0^1 d z ~ z^{N-1} \Delta^{\rm sv}_{ab}(z)
	\equiv g_0 \exp\left(\Psi^{sv}_{N}\right),
\end{align}
where $g_0$ is independent of the Mellin variable $N$, and the exponent $\Psi^{sv}_N$ resums the large logarithms $\ln^i N$. Up to next-to-next-to-leading logarithmic (NNLL) accuracy, $\Psi^{sv}_N$ takes the form:
\begin{align}\label{eq:gn}
	\Psi^{sv}_N = \ln(\bar{N})\,{g}_1(\bar{N}) + {g}_2(\bar{N}) + a_{s} \,{g}_3(\bar{N}) + \cdots,
\end{align}
with $\bar{N} = N\exp^{\gamma_E}$.
The universal functions ${g}_i$ are well known, and their expressions can be found in~\cite{Moch:2005ba, Catani:2003zt, Banerjee:2018vvb}. 
The constant $g_0$ contains the N-independent contributions and can be written as a series expansion in the strong coupling constant $a_{s}\left(g_s^2/(16 \pi^2)\right)$:
\begin{align}\label{eq:g0}
	g_0 = 1 + a_s \,g_{01} + a_{s}^{2} \,g_{02} + \cdots,
\end{align}
The coefficients $g_{01}$ and $g_{02}$ are process-dependent and are obtained following the approach described in~\cite{Ahmed:2020nci}. The evaluation of the $g_{02}$ coefficient, which is required for NNLL accuracy, involves the one-loop, one-loop squared, and two-loop amplitudes. The one-loop and one-loop squared amplitudes are computed using in-house codes. For the two-loop contribution, we make use of the public package \texttt{VVamp}~\cite{Gehrmann:2015ora} to construct the finite amplitudes, which are then used in the expression for $g_{0}$.

The resummed cross-section in $z$-space can now be obtained by performing the inverse Mellin transformation of the resummed cross-section in Mellin space:
\begin{align}
	\frac{d\sigma^{\text{N}^n\text{LL}}}{d Q} =
	\frac{d\hat{\sigma}^{(0)}}{d Q}
	\sum_{a,b \in \{q,\bar{q}\}}
	\int_{c-i\infty}^{c+i\infty}
	\frac{dN}{2\pi i}
	\tau^{-N}
	f_{a,N}(\mu_F)\,
	f_{b,N}(\mu_F)\,
	\hat{\sigma}_N^{\text{N}^n\text{LL}} \,.
\end{align}

Here, following the minimal prescription~\cite{Catani:1996yz}, we choose the contour of integration in the complex plane to avoid the Landau pole at $N = \exp(1/(2 a_{s} \beta_0) - \gamma_E)$. We set $N = c + x\exp^{i\phi}$ with $c = 1.9$ and $\phi = 3\pi/4$, as in~\cite{Vogt:2004ns}.

The differential cross-section after matching to the fixed-order counterparts is given by:
\begin{align}\label{eq:matched}
	\frac{d\sigma^{\text{N}^n\text{LO}+\text{N}^n\text{LL}}}{d Q} &=
	\frac{d\sigma^{\text{N}^n\text{LO}}}{d Q} +
	\frac{d\hat{\sigma}^{(0)}}{d Q}
	\sum_{a,b \in \{q,\bar{q}\}}
	\int_{c-i\infty}^{c+i\infty}
	\frac{dN}{2\pi i}
	\tau^{-N}
	f_{a,N}(\mu_F)\,
	f_{b,N}(\mu_F) \nonumber \\
	&\times
	\left[
	\hat{\sigma}_N^{\text{N}^n\text{LL}} -
	\left.\hat{\sigma}_N^{\text{N}^n\text{LL}}\right|_{\text{tr}}
	\right].
\end{align}
Here, $\left.\hat{\sigma}_N^{\text{N}^n\text{LL}}\right|_{\text{tr}}$ is the truncated resummed cross-section at the $n^{\text{th}}$ order to avoid double counting. 
Mellin-space PDFs, $f_{a,N}$, are obtained via \texttt{QCD-PEGASUS}~\cite{Vogt:2004ns}.

\section{Numerical Results}

We present the numerical results for the on-shell production of the $Z$-boson as well as $W$-boson pairs at the LHC for $\sqrt{S}=13.6$ TeV.
We present the results for the invariant mass distribution and total production cross section. For both the observable, the choice of unphysical factorisation and renormalisation scales is taken to be $\mu_R=\mu_F=\mu_0$. Here $\mu_0$ is the central scale, which by default is chosen to be the invariant mass $Q$ of the final state.

The input parameters are taken as follows: The masses of the $Z$ and $W$-bosons are taken as $m_{Z}=91.1876 $ GeV and $m_{W}=80.385$ GeV, respectively. 
The Weinberg angle is taken as $\text{sin}^2\theta_\text{w} = (1 - m_w^2/m_{z}^2) = 0.222897$ and the fine structure constant is $\alpha \simeq 1/132.2332$ that correspond to $G_F = 1.166379\times 10^{-5} \text{ GeV}^{-2}$.
We consider the five-flavour scheme (5FS) for $ZZ$ production, whereas for $WW$ production we choose the four-flavour scheme (4FS).
The choice of PDFs with the 5FS, we take MSHT20lo\_as130, MSHT20nlo\_as120 and MSHT20nnlo\_as118 PDF sets ~\cite{Bailey:2020ooq}.
For the PDFs with $n_f=4$, we take NNPDF30\_lo\_as\_0118\_nf\_4, NNPDF30\_nlo\_as\_0118\_nf\_4 and NNPDF30\_nnlo\_as\_0118\_nf\_4 PDF \cite{NNPDF:2017mvq}, all of them are taken from the {\tt LHAPDF} \cite{Buckley:2014ana}, using the central set (iset=0) default choice.
For $WW$ production, the 4FS is chosen because it avoids resonant top-quark contamination by removing bottom-quark emission sub-processes \cite{Gehrmann:2014fva}. 
The strong coupling constant at $m_{z} = 91.1876$ GeV is obtained from the respective PDF grids and then evolved to the desired scale using routines from {\tt LHAPDF}.
The NLO and the NNLO fixed-order results are taken from the 
public package {\tt MATRIX} \cite{Grazzini:2017mhc}.
Moreover, at the NLO level, these results are compared with those obtained from MadGraph \cite{Alwall:2011uj} and are found to agree well. 
For the resummation, we have completely used in-house developed numerical codes with the two loop amplitudes taken from the {\tt VVamp} package \cite{Gehrmann:2015ora}.
We define the following K-factors to quantify the size of the higher-order QCD corrections:
\begin{align}
& K_{\text{mn}}
=
	\frac{\sigma_{\text{N}^m\text{LO}}}{\sigma_{\text{N}^n\text{LO}}}
\,,~
R_{\text{mn}}
=
\frac{\sigma_{\text{N}^m\text{LO} + \text{N}^m\text{LL}}}{\sigma_{\text{N}^n\text{LO}}}
\label{eq:ratio}
\end{align}

\begin{figure}[ht!]
	\centerline{
		\includegraphics[scale =0.37]{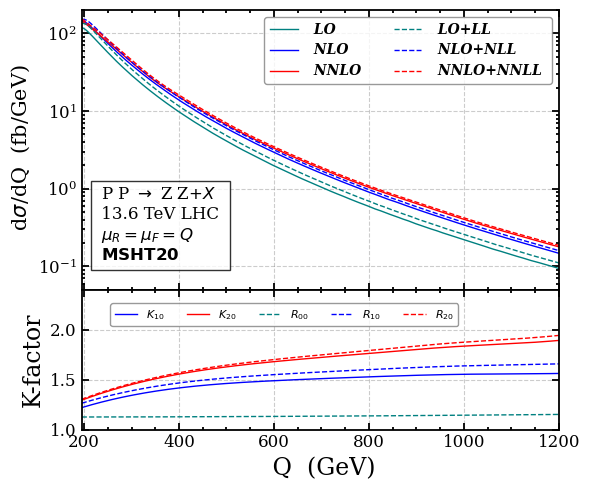}
		\includegraphics[scale =0.37]{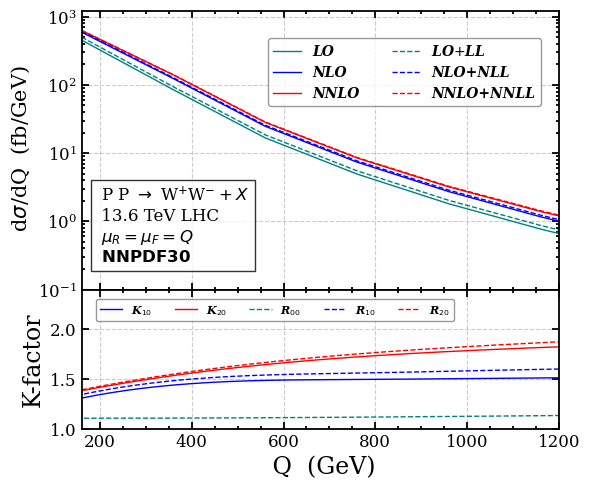}
	}
	\vspace{-2mm}
	\caption{\small{Resummed predictions for the invariant mass distribution of the  VV(V=Z,W) production and the corresponding K-factors up to NNLO+NNLL.}}
	\label{fig:match_inv}
\end{figure}

In figure \ref{fig:match_inv}, we present the invariant mass distribution for the $ZZ$ and $WW$ production processes at the LHC. 
The resummed predictions are shown up to NNLO+NNLL accuracy, along with the corresponding K-factors as defined in Eq.\ref{eq:ratio}.
We observe that including NNLL resummation increases the fixed-order NNLO results by a few per cent across the invariant-mass distribution for both $ZZ$ and $WW$ production. 
We notice that for WW production, the NNLO K-factor ($K_{20}$) varies from $1.38$ to $1.81$ over the range considered here. 
Whereas the NNLO+NNLL K-factor ($R_{20}$) changes from $1.39$ to $1.86$. 
The corrections coming from the NNLL resummation are more pronounced in the high invariant mass region. 
This is expected as the threshold logarithms become more significant in this region, leading to larger corrections from the resummation. 
We observe similar behaviour in $ZZ$ production as well.

\begin{figure}[ht!]
	\centerline{
		\includegraphics[scale =0.29]{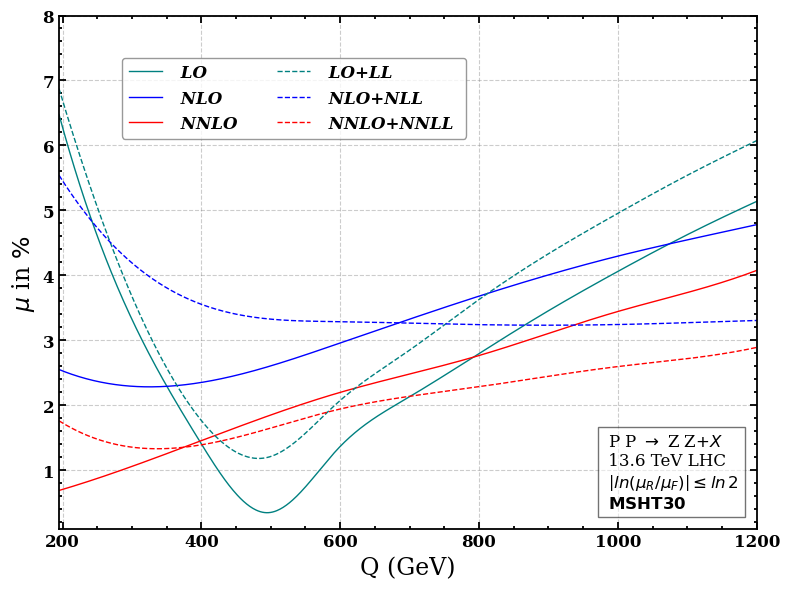}
		\includegraphics[scale =0.29]{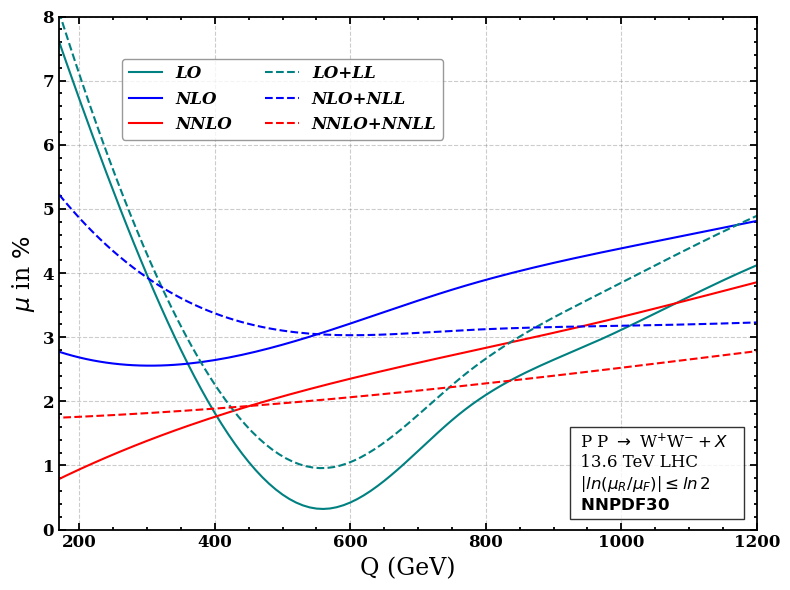}
	}
	\vspace{-2mm}
	\caption{\small{Seven-point scale uncertainties for VV production up to NNLO+NNLL.}}
	\label{fig:match_mu}
\end{figure}

In figure \ref{fig:match_mu}, we present the seven-point scale uncertainties for the invariant mass distribution of the $ZZ$ and $WW$ production processes at the LHC. 
We observe that including NNLL resummation significantly reduces scale uncertainties compared to fixed-order NNLO results.
For the $ZZ$ invariant mass distribution, the 7-point scale uncertainties decrease from $4.06\%$ at NNLO to $2.88\%$ at NNLO+NNLL, for $Q=1200$ GeV. 
Whereas for the $WW$ invariant mass distribution, the 7-point scale uncertainties decrease from $3.74\%$ at NNLO to $2.72\%$ at NNLO+NNLL, at the same $Q$ value. 
This reduction in scale uncertainties is a clear indication of the improved perturbative stability of the predictions in the high-Q region, when the NNLL resummation is included.

\begin{table}[ht!]
\begin{center}
\begin{tabular}{|c|cc|cc|cc|}
\hline
\multirow{2}{*}{Q(GeV)} & \multicolumn{2}{c|}{$\mu_{0}$ = $2$Q}                               & \multicolumn{2}{c|}{$\mu_{0}$ = Q}                                & \multicolumn{2}{c|}{$\mu_{0}$ = Q/$2$} \\  \cline{2-7}
                        & \multicolumn{1}{c|}{${\text{NNLO}}$} & ${\text{NNLO+NNLL}}$          & \multicolumn{1}{c|}{${\text{NNLO}}$} & ${\text{NNLO+NNLL}}$           & \multicolumn{1}{c|}{${\text{NNLO}}$}   & ${\text{NNLO+NNLL}}$  \\ \hline
165                     & \multicolumn{1}{c|}{0.71} & 1.37                                     & \multicolumn{1}{c|}{0.74} & 1.74                                                 & \multicolumn{1}{c|}{0.99}   & 2.14       \\ \hline
1365                    & \multicolumn{1}{c|}{3.97} & 2.80                                     & \multicolumn{1}{c|}{4.29} & 2.99                                                 & \multicolumn{1}{c|}{4.49}   & 3.28       \\ \hline
1765                    & \multicolumn{1}{c|}{4.76} & 3.19                                     & \multicolumn{1}{c|}{5.22} & 3.43                                                 & \multicolumn{1}{c|}{5.48}   & 3.72       \\ \hline
\end{tabular}
\caption{The 7-point scale uncertainty (in $\%$) for the different central scale choices at different Q values for WW production.}
\label{tab:tableWW_Q}
\end{center}
\end{table}
\vspace{-0.8cm}
\begin{figure}[ht!]
	\centerline{
		\includegraphics[scale =0.43]{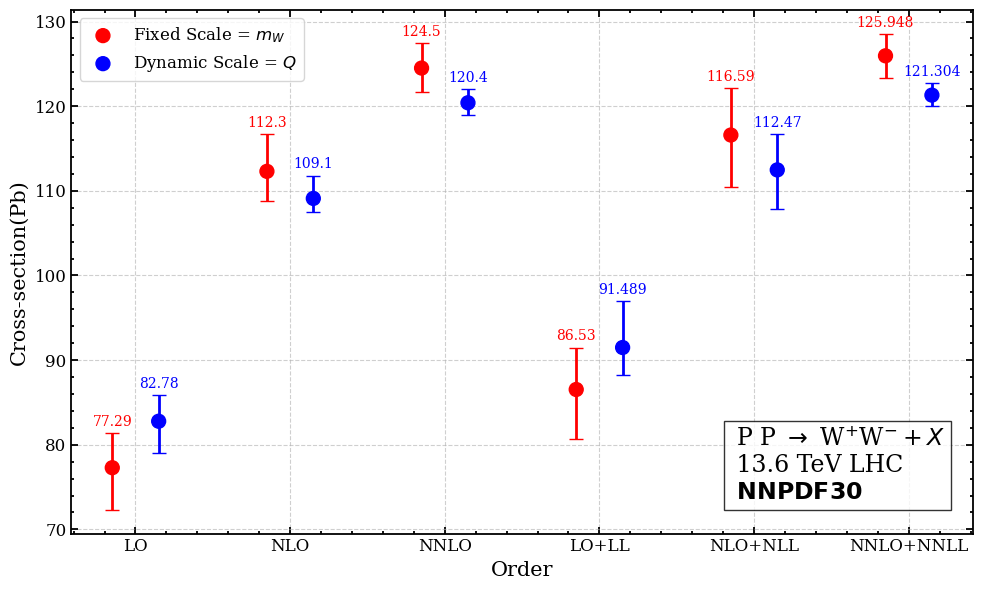}
	}
	\caption{\small{Resummed predictions for the invariant mass distribution of the  VV(V=Z,W) production and the corresponding K-factors up to NNLO+NNLL.}}
	\label{fig:fixed_dynamic}
\end{figure}
We also present the 7-point scale uncertainties for the total cross section of the $WW$ production process at different central scale choices in Table \ref{tab:tableWW_Q}. 
We find that the inclusion of NNLL resummation leads to an increase in the scale uncertainties for the total production, which is expected, as in the low Q region (where most of the cross-section comes from) of the invariant mass distribution, the resummed corrections increase the scale uncertainties compared to the fixed order results. 
We also notice that the scale uncertainties are generally smaller for the central scale choice $\mu_0 =2Q$ compared to $\mu_0 = Q$ and $\mu_0 = Q/2$.
In figure \ref{fig:fixed_dynamic}, we study the total production cross section for the $WW$ production with fixed central scale choice $\mu_0 = m_{W}$ and dynamic central scale choice $\mu_0 = Q$. 
We find that the fixed-scale choice leads to larger scale uncertainties than the dynamic-scale choice for all orders, both in fixed-order and resummed predictions. 

\section{Conclusion}

We have presented threshold resummation for the VV production processes. We have performed the resummation at NNLO+NNLL accuracy.
The contribution of NNLL resummed results is found to increase the fixed order NNLO results by a few per cent in the invariant mass distribution as well as in the total cross section.
We have found that the NNLO+NNLL resummed results exhibit less scale uncertainty in the high-Q region than the NNLO results.
For the ZZ invariant mass distribution, 7-point scale uncertainties decrease from $4.06\%$ at NNLO to $2.88\%$ at NNLO+NNLL, at  $Q=1200$GeV. Whereas for the WW invariant mass distribution, the 7-point scale uncertainties decrease from $3.74\%$ at NNLO to $2.72\%$ at NNLO+NNLL at $ Q=1200$ GeV. 
In addition, we analyse the seven-point scale uncertainties by varying the central scale in the range $Q/2$ to $2Q$ for WW production.
The results show that the choice $2 Q$ yields smaller uncertainties at both the fixed-order and the resummed predictions.
Alongside this, we study the effect of the fixed central scale choice compared to the dynamic scale, and we find that the dynamic scale choice yields much more stable results than the fixed one.
The NNLO+NNLL results presented should help improve precision studies at the LHC and are also expected to be useful for future hadron collider analyses.

\end{document}